\documentclass[aps,prl,twocolumn]{revtex4}
\usepackage{amssymb}
\usepackage{graphicx}

\begin{document}

\title{Peripheral mixing of passive scalar at small Reynolds number}

\author{G. Boffetta, F. De Lillo}
\affiliation{Dipartimento di Fisica Generale and INFN, Universit\`a di Torino,
via P.Giuria 1, 10125 Torino (Italy) \\
and CNR-ISAC, Sezione di Torino, corso Fiume 4, 10133 Torino (Italy)}
\author{A. Mazzino}
\affiliation{Dipartimento di Fisica, Universit\`a di Genova, INFN and CNISM,
via Dodecaneso 33, 16146 Genova (Italy)}

\date{\today}

\begin{abstract}
% insert abstract here
Mixing of a passive scalar in the peripheral region close to 
a wall is investigated by means of accurate direct numerical simulations
of both a three-dimensional Couette channel flow at low Reynolds
numbers and a two-dimensional synthetic flow.  In both cases, the resulting
phenomenology can be understood in terms of the theory recently 
developed by Lebedev and Turitsyn [Phys. Rev. E 69, 036301, 2004]. 
Our results prove the robustness of the identified mechanisms responsible 
for the persistency of scalar concentration close to
the wall with important consequences in completely different fields 
ranging from microfluidic applications to environmental dispersion modeling.
\end{abstract}

\pacs{PACS?}
%\keywords{}

\maketitle

Mixing of passive tracer in a turbulent flow is a fundamental problem
of great technological interest which has undergone a significant 
theoretical progress in the last years \cite{ss_nature00,fgv_rmp01}. 
At low diffusivity and small scales the viscous-convective Batchelor 
regime of mixing arises for a large value of the Schmidt number 
$Sc=\nu/\kappa$, where $\nu$ is the kinematic viscosity and $\kappa$ 
the molecular diffusivity \cite{batchelor_jfm59}. 
In this regime the velocity field can be
considered smooth because of the exponential decay of the velocity
spectrum. An analogous situation is realized if the tracer is advected
by a time dependent, chaotic flow, i.e. a spatially smooth flow at
all scales.
Several theoretical predictions, including the exponential decay in 
time of passive scalar fluctuations 
\cite{pierrehumbert_csf94,son_pre99,bf_pre99,hv_pof05},
have been verified in experiments and
numerical simulations \cite{gs_nature01,wmg_pof97}.

Recently, it has been shown that the presence of boundaries can alter
significantly the predictions based on an unbounded domain.
No slip boundary condition for the velocity field reduces the 
efficiency of mixing close to the boundary -- the peripheral
region -- which becomes a source of passive scalar. 
This effect is of particular importance in the case of boundary
dominated geometries, such as in microfluidics. 
Indeed, it is well known that the lack of efficient mixing is one 
of the problems in many microfluidic devices which operate at
vanishing Reynolds number \cite{sq_rmp05}.
Several statistical predictions for the peripheral mixing
have been recently made \cite{cl_prl03,lt_pre04} and checked with
laboratory experiments in a chaotic microchannel \cite{sg_prl05},
in polymer solutions \cite{bss_prl06} and in kinematic
simulations \cite{sh_pof07}.

\begin{figure}
\includegraphics[clip=true,keepaspectratio,width=8.0cm]{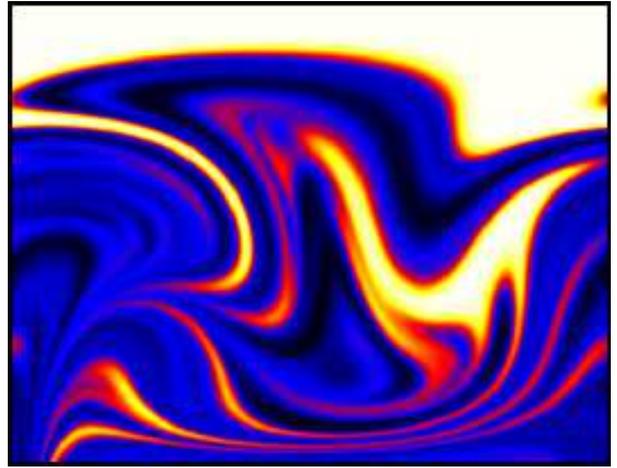}
\caption{Snapshot of passive tracer concentration close to a wall
placed on the top. White correspond to high concentration, black to
low concentration. The initial condition is a step function in the 
vertical direction. Tracers is advected according to (\ref{eq:1}) with
a two-dimensional synthetic velocity field.}
\label{fig1}
\end{figure}

In this letter, we will discuss the problem of tracer mixing 
in presence of boundaries by means of direct numerical simulation of 
a Couette flow at small Reynolds number and in kinematic simulations
of a chaotic flow.
Before discussing quantitative results,
we can have a physical intuition from 
Figure~\ref{fig1} which shows a snapshot of tracer concentration
advected by a two-dimensional chaotic flow.
Because of the vanishing velocity, the tracer persists for very long time 
close to the boundary from which it is only intermittently transported into
the bulk in the form of elongated filaments. 

%------------------------------------------------------------------------
%------------------------------------------------------------------------

A passive scalar $\theta$ advected by an incompressible velocity 
field ${\bf u}$ obeys the equation
\begin{equation}
\frac{\partial \theta}{\partial  t} + {\bf u}\cdot\nabla\theta=
\kappa\nabla^2\theta
\label{eq:1}
\end{equation}
where $\kappa$ is the molecular diffusivity, and appropriate initial 
and boundary conditions are set for $\theta$. 
In general ${\bf u}$ is subject to its own set of equations,
the typical case being Navier-Stokes equations, together with boundary
conditions and stirring mechanism which therefore determine the precise 
form of ${\bf u}$ close to the boundary.
However some general predictions on the evolution of $\theta$ can be 
made on the basis of the no-slip conditions and incompressibility only,
assuming the flow to have a short correlation time with respect to the 
typical mixing time. 
To simplify the notation, but without loss of generality, we assume the 
flow to be two-dimensional, with $(x,y)$ the coordinates parallel and
normal to the wall ($y=0$ corresponding to the wall) 
and $(u,v)$ the associated velocity components.

As a consequence of no slip (${\bf u}(x,y=0)=0$) and incompressibility
($\nabla\cdot{\bf u}=0$) conditions, there is a region close to the wall, 
characterized by the scaling $\langle u^2(y) \rangle \sim y^2$ and
$\langle v^2(y) \rangle \sim y^4$ (where by $<\cdot>$ we indicate the
double average with respect to the coordinates parallel to the wall and
velocity realizations). The velocities in the bulk of the container or 
duct are therefore much more intense with respect to the ones in the 
layers close to the wall, and the passive scalar evolution becomes faster
and faster as one moves away from the wall.
Starting from these considerations, it is possible to describe the 
evolution of a passive tracer initially concentrated in a layer of
thickness $\delta$ close to the wall in term of a turbulent diffusivity.
Averaging (\ref{eq:1}), the equation for the $y$-evolution of the scalar
profile is
\begin{equation}
\partial_t\langle\theta\rangle=\mu\partial_y 
[ y^4\partial_y\langle\theta\rangle ] 
+ \kappa\partial^2_y\langle\theta\rangle. \,.
\label{eq:2}
\end{equation}

The first term in the rhs of (\ref{eq:2}) describes the role of chaotic 
advection in terms of an eddy diffusivity
$D=\int_0^\infty\langle v(y,0)v(y,t)\rangle dt \sim \mu y^4$.
Comparison with the second term in the rhs of (\ref{eq:2}) suggests 
that the evolution of the 
profile is dominated by advection as long as 
$\delta\gtrsim r_{bl}=(\kappa/\mu)^{1/4}$, the thickness of the 
diffusive boundary layer.
Under this condition, the diffusive contribution can be neglected 
and (\ref{eq:2}) becomes
\begin{equation}
\partial_t\langle\theta\rangle=\mu\partial_y y^4\partial_y\langle\theta\rangle 
\label{eq:3}
\end{equation}

Taking as initial condition for the scalar a distribution concentrated
at the wall with $\theta(x,0;0)=1$ and 
$lim_{y\rightarrow\infty}\theta(x,y;t)=0$, the asymptotic solution 
of (\ref{eq:3}) for large times is \cite{lt_pre04}
\begin{equation}
\langle\theta(y,t)\rangle=
\left[ \mathrm{erf}\left(\frac{\delta}{2y}\right)-
\frac{\delta}{\sqrt{\pi}y}\exp\left(-\frac{\delta^2}{4y^2}\right)\right]
\label{eq:4}
\end{equation}
i.e. the profile has a universal form, independently on the details of
the initial distribution. The thickness $\delta=(\mu t)^{-1/2}$ is the 
only characteristic scale and decreases in time, as the layer occupied
by the scalar contracts in the evolution.
It is important to remark that the specific form of the profile
(\ref{eq:4}) gives a practically constant 
concentration for $y\lesssim \delta/4$, making the boundary conditions 
for the scalar irrelevant in the advective stage.
We remark that although (\ref{eq:1}) obviously conserves the average scalar
$\langle \theta \rangle$, from (\ref{eq:4}) one has that 
$\int \langle \theta(y,t) \rangle dy = \delta(t)/\sqrt{\pi}\simeq t^{-1/2}$
is time dependent. The reason is that in deriving (\ref{eq:4})
the bulk is considered as an infinite reservoir for the scalar 
which has therefore zero average.

Neglecting diffusion, the advection equation (\ref{eq:1}) holds 
for any local function of $\theta$ too and therefore its average
is governed by a generalization of (\ref{eq:3}). In particular the 
moments of scalar concentration $\langle \theta^n \rangle$ are 
expected to follow the same profile (\ref{eq:4}) independently 
of $n$. 
This is the mathematical consequence of the intermittent
nature of scalar advection shown in Figure~\ref{fig1} in which 
$\langle \theta^n \rangle$ is dominated by the white regions in which
$\theta=1$.
It is also possible to derive 
an expression for the probability density function valid for
any time and distance from the wall as \cite{lt_pre04}
\begin{eqnarray}
P(\theta,y,t)&=&\frac{1}{y y_0 |\theta_0'(y_0)|}
\left[(1-2 \mu t y y_0) g\left({1 \over y}-{1 \over y_0}\right) \right. 
\nonumber \\
&& \left. +(1+2 \mu t y y_0) g\left({1 \over y}+{1 \over y_0}\right) \right]
\label{eq:5}
\end{eqnarray}
where $g(x)$ is a Gaussian distribution with zero mean and variance
$2 \mu t$ and
the dependence on $\theta$ is defined implicitly through the
monotonic, homogeneous along the wall, initial profile:
$\theta_0(y_0)=\theta$ (note that large $\theta$ correspond to small
$y_0$). 
At variance with (\ref{eq:4}), which is valid only asymptotically in time, 
the prediction for the PDF is valid for any times as it depends on the 
details of the initial distribution. It is interesting to observe that
(\ref{eq:5}) predicts that $P(\theta=1,y,t)=0$ for any $y>0$: in the absence
of diffusion it takes an infinite time for the scalar to be transported 
away from the wall.

We have tested the above theoretical predictions
by means of numerical simulations
of a three-dimensional Couette flow and a synthetic two-dimensional
flow. 
For the first case, the Navier-Stokes equations have been   numerically integrated
in a 3D slab geometry of dimension $L_x \times 2 L_y \times L_z$
with no-slip boundary conditions on the 
two planes $y=0$ and $y=2 L_y$, and periodic boundary conditions on the
streamwise and spanwise directions $x$ and $z$.
The flow is forced by the relative motion of the two opposed walls
with opposite velocity $\pm U_0$,
from which a large scale Reynolds number is defined as
$Re=U_0 L_y/\nu$.
Direct numerical simulations are performed by means of a standard 
pseudo-spectral Fourier-Chebyshev code \cite{chqz_88,de_pre07} at resolution
$128\times 65\times 128$ for a domain of size 
$L_x\times 2 L_y\times L_z=8\times 2\times 8$.
We use a moderate Reynolds number $Re \simeq 600$,
which is sufficiently large to sustain a turbulent-like motion for
long time \cite{mmg_pof95} but still small in order to have a well developed
viscous layer where scaling imposed by boundaries is observed.
The average wall shear rescaled with mean shear is 
$s_0/(U_0/L_y) \simeq 3.3$, the rescaled friction velocity 
$u_*/U_0=\sqrt{\nu s_0}/U_0 \simeq 0.074$ and the friction Reynolds number 
$Re^*\simeq 44$. Scales and times are made dimensionless with
the half-channel height $L_y$ and large scale time $T=L_y/U_0$.

%------------------------------------------------------------------------
\begin{figure}[htb]
\includegraphics[clip=true,keepaspectratio,width=8.0cm]{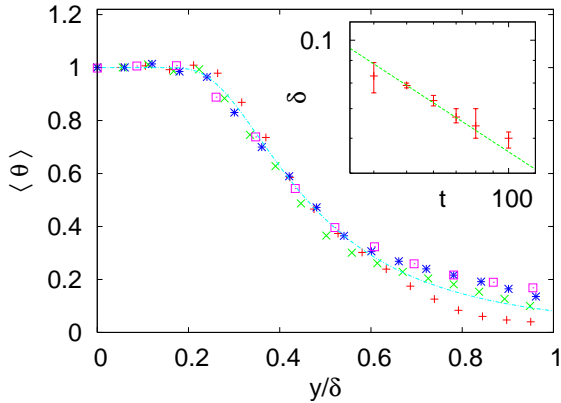}
\caption{Profile of the scalar density $\theta$ at different times
in the Couette channel rescaled with respect to $\delta(t)$ and compared 
with the theoretical prediction (\ref{eq:4}) (continuous line). 
The inset shows the fitted values of $\delta$ at different times
and the prediction $\delta \simeq t^{-1/2}$.}
\label{fig2}
\end{figure}
%------------------------------------------------------------------------

In order to avoid the effects of diffusivity and increase the available 
observation range, the simulation of the tracer advection (\ref{eq:1})
was carried out with a Lagrangian method: trajectories of $N=10^7$
particles, representing the concentration of tracer,
are integrated according to the equation
$\dot{\mathbf{x}}(t)=\mathbf{u}(\mathbf{x},t)$.         
The initial condition for the particles is an uniform distribution 
in the $x$ and $z$ directions in a layer close to the wall 
$y \le 0.04 L_y$.
Continuous tracer distribution $\theta({\bf x},t)$ is reconstructed
at every time proportional to local tracer density on the grid.
The advantage of the present method is the possibility to perform
the simulation at Schmidt number virtually infinite (although
a small numerical diffusion is always present), at the price 
of some noise in the reconstruction of small scales due to the 
discreteness of the tracer.

In Fig.~\ref{fig2} we plot the mean profile $\langle \theta(y,t) \rangle$
at different times compared with the theoretical curve (\ref{eq:4}).
The value of the thickness $\delta$ at different time is obtained 
from the fit of the profiles with (\ref{eq:4}) and its dependence
on $t$ is compatible with the prediction $\delta=(\mu t)^{-1/2}$,
from which $\mu \simeq 3.2$ is estimated. 

Figure~\ref{fig2} shows that scalar profiles in the bulk deviate 
from the theoretical curve at large $y$. This is because,
even if the value of $Re$ is small, the $y^4$ scaling characteristic 
of the viscous layer, can be observed 
only up to a distance $y\sim0.1 L_y$ from the wall. 
Therefore, in order to extend the range of scaling,
we performed an additional set of simulations based on a kinematic 
model for the velocity field.
We define a two-dimensional velocity field in terms of a synthetic
stream function $\Psi(x,y)=\Phi(x,y)G(y)$, where 
$\Phi=\sin(k_x x+\varphi_x(t))\sin(k_y y+\varphi_y(t))$ represents a
time-dependent cellular flow while $G(y)$ is tailored to
reproduce the correct scaling at the wall. 
If $G\sim y^2$ for $y$ close to the wall, the scaling of both
components of the velocity is ensured. 
We chose $G(y)\simeq [1-\cos(k_s y))]$ close to each wall,
with $k_s=\pi/(2L_s)$ and the position $L_s$ of the inflection 
point defining the width of the scaling region
for the velocity, $y\lesssim L_s$, which is the region of interest. 
In the bulk, a matching function connects the profiles.
The phases of the cellular flow, $\varphi_x$ and $\varphi_y$ are given by
a random process with a finite correlation time. 
The velocity field generated by $\Phi$ is placed on a grid of
size $L_x=\pi$ and $2 L_y=4 \pi$ at resolution $512 \times 2048$ where 
the evolution of (\ref{eq:1}) is integrated by means of a pseudo-spectral
code, with periodic boundary conditions. 
This numerical approach is similar to that used in \cite{sh_pof07,cl_jetp08}. 
Scaling regions extend approximatively to $L_s=4$.
The time unit is chosen so that the correlation time of the velocity field 
is $T=1$. In these units we have $\mu\simeq 2.66$ and 
$\kappa=3.42\times 10^{-6}$ and therefore the width of the diffusive boundary
layer is $r_{bl}\simeq 0.034 L_y$.
As initial condition we choose a distribution null in the bulk and 
concentrated at the walls in two smoothed-step functions of size $L_y/4$. The
results are based on ensemble average over $100$
realizations of the random noise driving the kinematic velocity field. 

%------------------------------------------------------------------------
\begin{figure}
\includegraphics[clip=true,keepaspectratio,width=8.0cm]{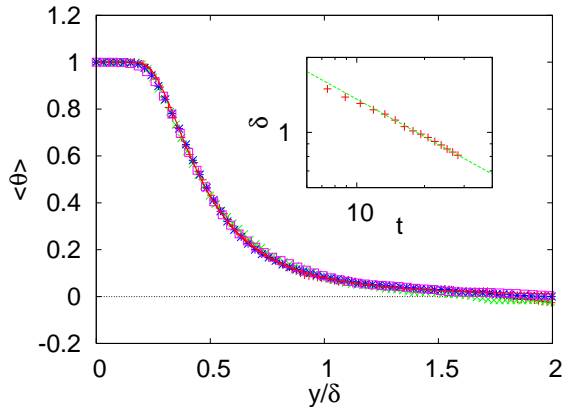}
\caption{Average profile of the scalar density $\theta$ at different times
rescaled with respect to $\delta$ and the bulk average $b$ (see text)
and compared with (\ref{eq:4}) (continuous line).
In the inset, the thickness of the profile is shown together with
the prediction $t^{-1/2}$ dependence.}
\label{fig3}
\end{figure}
%------------------------------------------------------------------------

In Fig.~\ref{fig3} the profile $\langle \theta(y,t) \rangle$ 
is compared, at different times, with the theoretical curve. 
In order to  accurately resolve the region close to the wall,
the extension of the domain in the bulk is not large enough for
the approximation of an infinite bulk to be valid. 
Therefore, because of the conservation of $\langle \theta \rangle$,
after a short transient a
relevant amount of scalar accumulates in the central region of the domain, 
thus affecting the overall shape of the profile. 
In order to compare the numerical
results with the theoretical prediction based on an infinite basin,
we computed the profile of 
the auxiliary field $\tilde{\theta}=(\theta-b)/(1-b)$, where $b(t)$ 
is the time-dependent value of $\theta$ in the bulk (averaged over $x$).
Figure~\ref{fig3} shows the remarkable agreement obtained between 
theory and numerics, indicating
that the profile (\ref{eq:4}) can be easily extended to the general case of
advection in a finite vessel. From the fitting procedure we get the values of
the parameter $\delta$, which is found to follow accurately the prediction
$\delta(t)=(\mu t)^{-1/2}$ with $\mu \simeq 2.13$
(see inset of Fig.~\ref{fig3}). 

\begin{figure}
\includegraphics[clip=true,keepaspectratio,width=8.0cm]{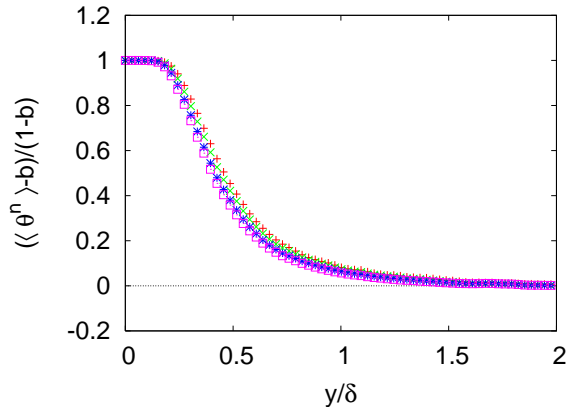}
\caption{Average profiles of moments of scalar density
$\langle\theta^n\rangle$ for $n=1$ ($+$), $n=2$ ($\times$), 
$n=4$ ($*$) and $n=6$ ($\Box$) at time $t=18 T$ in the two-dimensional 
kinematic simulation.}
\label{fig4}
\end{figure}

Figure~\ref{fig4} shows the profiles of different moments of 
scalar concentration $\langle \theta^n(y,t) \rangle$ computed at an 
intermediate time. All the moments collapse on the prediction
(\ref{eq:4}), confirming the fact that in this stage diffusion
is negligible and mixing of the scalar is dominated by 
eddy diffusivity according to (\ref{eq:3}).

%------------------------------------------------------------------------
%------------------------------------------------------------------------

\begin{figure}
\includegraphics[clip=true,keepaspectratio,width=7.0cm]{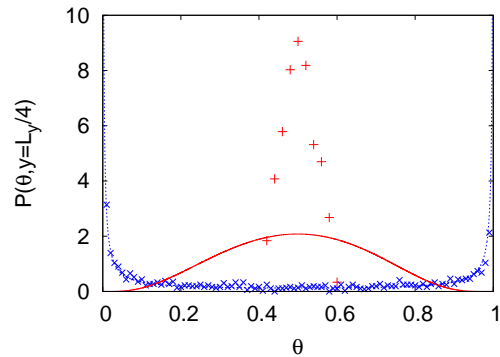}
\caption{PDF's of the scalar density in the kinematic simulation at $y=L_y/4$,
corresponding to the thickness of the initial distribution at times
$t=0.02$ ($+$) and $t=3.0$ ($\times$). Lines show theoretical predictions.}
\label{fig5}
\end{figure}

We have also computed the probability density function, $P(\theta,t,y)$, 
of the passive scalar.  The results, at a distance $y=L_y/4$ from the 
wall, corresponding to the thickness of the initial distribution,
are shown in Fig.~\ref{fig5} for two different times.
At very short time the PDF is peaked around the initial condition 
$\vartheta_0/2$. Observe that the distribution is much more narrow
than that theoretically predicted by (\ref{eq:5}), because the
eddy diffusivity approximation (\ref{eq:2}) is not justified at short
times. 

At longer time the PDF forms two pronounced peaks at the extreme 
values which are responsible for the saturation of the moments of $\theta$, 
since the peak in $\theta=1$ gives a dominant contribution to any 
$\langle \theta^n\rangle$.
Physically the presence of the two peaks is related to long tongues 
protruding from the wall region into the bulk. 
Advection stretches such structures, while
preserving the value of $\theta$, the smoothing effect of molecular
diffusivity becoming effective only on longer times. 

%------------------------------------------------------------------------
%------------------------------------------------------------------------

In conclusion, our results for both the three-dimensional channel flow 
and the two-dimensional synthetic  flow  are accurately explained by
the theoretical description of Ref.~\cite{lt_pre04}.  Our results emphasize the 
importance of a correct description of the close-to-the-wall region 
where a scalar field tends to persist. 
The incorrect reproduction of near-wall scaling behavior 
$u(y)\propto  y$ and $v(y)\propto  y^2$ necessarily destroys the above 
mechanism. 
This might be a serious problem in applications of environmental 
dispersion modeling under strong stability conditions where the viscous 
layer can reach values of orders of meters thus affecting
the realm of human activities. A poor (in term of resolution) description 
of this region would result in a dangerous underestimation of the level 
of pollutants concentration close to the ground.

\acknowledgements
This work was supported by Piedmont Industrial Research Grant INUMICRO.

\bibliography{bibliografia}{}

\begin{thebibliography}{19}
\expandafter\ifx\csname natexlab\endcsname\relax\def\natexlab#1{#1}\fi
\expandafter\ifx\csname bibnamefont\endcsname\relax
  \def\bibnamefont#1{#1}\fi
\expandafter\ifx\csname bibfnamefont\endcsname\relax
  \def\bibfnamefont#1{#1}\fi
\expandafter\ifx\csname citenamefont\endcsname\relax
  \def\citenamefont#1{#1}\fi
\expandafter\ifx\csname url\endcsname\relax
  \def\url#1{\texttt{#1}}\fi
\expandafter\ifx\csname urlprefix\endcsname\relax\def\urlprefix{URL }\fi
\providecommand{\bibinfo}[2]{#2}
\providecommand{\eprint}[2][]{\url{#2}}

\bibitem[{\citenamefont{Shraiman and Siggia}(2000)}]{ss_nature00}
\bibinfo{author}{\bibfnamefont{B.}~\bibnamefont{Shraiman}} \bibnamefont{and}
  \bibinfo{author}{\bibfnamefont{E.}~\bibnamefont{Siggia}},
  \bibinfo{journal}{Nature} \textbf{\bibinfo{volume}{405}},
  \bibinfo{pages}{639} (\bibinfo{year}{2000}).

\bibitem[{\citenamefont{Falkovich et~al.}(2001)\citenamefont{Falkovich,
  Gaw\ifmmode~\mbox{\c{e}}\else \c{e}\fi{}dzki, and Vergassola}}]{fgv_rmp01}
\bibinfo{author}{\bibfnamefont{G.}~\bibnamefont{Falkovich}},
  \bibinfo{author}{\bibfnamefont{K.}~\bibnamefont{Gaw\ifmmode~\mbox{\c{e}}\else
  \c{e}\fi{}dzki}}, \bibnamefont{and}
  \bibinfo{author}{\bibfnamefont{M.}~\bibnamefont{Vergassola}},
  \bibinfo{journal}{Rev. Mod. Phys.} \textbf{\bibinfo{volume}{73}},
  \bibinfo{pages}{913} (\bibinfo{year}{2001}).

\bibitem[{\citenamefont{Batchelor}(1959)}]{batchelor_jfm59}
\bibinfo{author}{\bibfnamefont{G.~K.} \bibnamefont{Batchelor}},
  \bibinfo{journal}{J. Fluid Mech} \textbf{\bibinfo{volume}{5}},
  \bibinfo{pages}{113} (\bibinfo{year}{1959}).

\bibitem[{\citenamefont{Pierrehumbert}(1994)}]{pierrehumbert_csf94}
\bibinfo{author}{\bibfnamefont{R.~T.} \bibnamefont{Pierrehumbert}},
  \bibinfo{journal}{Chaos Solitons Fractals} \textbf{\bibinfo{volume}{4}},
  \bibinfo{pages}{1091} (\bibinfo{year}{1994}).

\bibitem[{\citenamefont{Son}(1999)}]{son_pre99}
\bibinfo{author}{\bibfnamefont{D.~T.} \bibnamefont{Son}},
  \bibinfo{journal}{Phys. Rev. E} \textbf{\bibinfo{volume}{59}},
  \bibinfo{pages}{R3811} (\bibinfo{year}{1999}).

\bibitem[{\citenamefont{Balkovsky and Fouxon}(1999)}]{bf_pre99}
\bibinfo{author}{\bibfnamefont{E.}~\bibnamefont{Balkovsky}} \bibnamefont{and}
  \bibinfo{author}{\bibfnamefont{A.}~\bibnamefont{Fouxon}},
  \bibinfo{journal}{Phys. Rev. E} \textbf{\bibinfo{volume}{60}},
  \bibinfo{pages}{4164} (\bibinfo{year}{1999}).

\bibitem[{\citenamefont{Haynes and Vanneste}(2005)}]{hv_pof05}
\bibinfo{author}{\bibfnamefont{P.}~\bibnamefont{Haynes}} \bibnamefont{and}
  \bibinfo{author}{\bibfnamefont{J.}~\bibnamefont{Vanneste}},
  \bibinfo{journal}{Physics of Fluids} \textbf{\bibinfo{volume}{17}},
  \bibinfo{pages}{097103} (\bibinfo{year}{2005}).

\bibitem[{\citenamefont{Groisman and Steinberg}(2001)}]{gs_nature01}
\bibinfo{author}{\bibfnamefont{A.}~\bibnamefont{Groisman}} \bibnamefont{and}
  \bibinfo{author}{\bibfnamefont{V.}~\bibnamefont{Steinberg}},
  \bibinfo{journal}{Nature} \textbf{\bibinfo{volume}{410}},
  \bibinfo{pages}{905} (\bibinfo{year}{2001}).

\bibitem[{\citenamefont{Williams et~al.}(1997)\citenamefont{Williams, Marteau,
  and Gollub}}]{wmg_pof97}
\bibinfo{author}{\bibfnamefont{B.}~\bibnamefont{Williams}},
  \bibinfo{author}{\bibfnamefont{D.}~\bibnamefont{Marteau}}, \bibnamefont{and}
  \bibinfo{author}{\bibfnamefont{J.}~\bibnamefont{Gollub}},
  \bibinfo{journal}{Physics of Fluids} \textbf{\bibinfo{volume}{9}}
  (\bibinfo{year}{1997}).

\bibitem[{\citenamefont{Squires and Quake}(2005)}]{sq_rmp05}
\bibinfo{author}{\bibfnamefont{T.~M.} \bibnamefont{Squires}} \bibnamefont{and}
  \bibinfo{author}{\bibfnamefont{S.~R.} \bibnamefont{Quake}},
  \bibinfo{journal}{Rev. Mod. Phys.} \textbf{\bibinfo{volume}{77}},
  \bibinfo{pages}{977} (\bibinfo{year}{2005}).

\bibitem[{\citenamefont{Chertkov and Lebedev}(2003)}]{cl_prl03}
\bibinfo{author}{\bibfnamefont{M.}~\bibnamefont{Chertkov}} \bibnamefont{and}
  \bibinfo{author}{\bibfnamefont{V.}~\bibnamefont{Lebedev}},
  \bibinfo{journal}{Phys. Rev. Lett.} \textbf{\bibinfo{volume}{90}},
  \bibinfo{pages}{034501} (\bibinfo{year}{2003}).

\bibitem[{\citenamefont{Lebedev and Turitsyn}(2004)}]{lt_pre04}
\bibinfo{author}{\bibfnamefont{V.~V.} \bibnamefont{Lebedev}} \bibnamefont{and}
  \bibinfo{author}{\bibfnamefont{K.~S.} \bibnamefont{Turitsyn}},
  \bibinfo{journal}{Phys. Rev. E} \textbf{\bibinfo{volume}{69}},
  \bibinfo{pages}{036301} (\bibinfo{year}{2004}).

\bibitem[{\citenamefont{Simonnet and Groisman}(2005)}]{sg_prl05}
\bibinfo{author}{\bibfnamefont{C.}~\bibnamefont{Simonnet}} \bibnamefont{and}
  \bibinfo{author}{\bibfnamefont{A.}~\bibnamefont{Groisman}},
  \bibinfo{journal}{Phys. Rev. Lett.} \textbf{\bibinfo{volume}{94}},
  \bibinfo{pages}{134501} (\bibinfo{year}{2005}).

\bibitem[{\citenamefont{Burghelea et~al.}(2006)\citenamefont{Burghelea, Segre,
  and Steinberg}}]{bss_prl06}
\bibinfo{author}{\bibfnamefont{T.}~\bibnamefont{Burghelea}},
  \bibinfo{author}{\bibfnamefont{E.}~\bibnamefont{Segre}}, \bibnamefont{and}
  \bibinfo{author}{\bibfnamefont{V.}~\bibnamefont{Steinberg}},
  \bibinfo{journal}{Phys. Rev. Lett.} \textbf{\bibinfo{volume}{96}},
  \bibinfo{pages}{214502} (\bibinfo{year}{2006}).

\bibitem[{\citenamefont{Salman and Haynes}(2007)}]{sh_pof07}
\bibinfo{author}{\bibfnamefont{H.}~\bibnamefont{Salman}} \bibnamefont{and}
  \bibinfo{author}{\bibfnamefont{P.~H.} \bibnamefont{Haynes}},
  \bibinfo{journal}{Phys. Fluids} \textbf{\bibinfo{volume}{19}},
  \bibinfo{pages}{067101} (\bibinfo{year}{2007}).

\bibitem[{\citenamefont{Canuto et~al.}(1988)\citenamefont{Canuto, Hussaini,
  Quarteroni, and Zang}}]{chqz_88}
\bibinfo{author}{\bibfnamefont{C.}~\bibnamefont{Canuto}},
  \bibinfo{author}{\bibfnamefont{M.}~\bibnamefont{Hussaini}},
  \bibinfo{author}{\bibfnamefont{A.}~\bibnamefont{Quarteroni}},
  \bibnamefont{and} \bibinfo{author}{\bibfnamefont{T.}~\bibnamefont{Zang}},
  \emph{\bibinfo{title}{{Spectral methods in fluid dynamics}}}
  (\bibinfo{publisher}{Springer-Verlag New York}, \bibinfo{year}{1988}).

\bibitem[{\citenamefont{De~Lillo and Eckhardt}(2007)}]{de_pre07}
\bibinfo{author}{\bibfnamefont{F.}~\bibnamefont{De~Lillo}} \bibnamefont{and}
  \bibinfo{author}{\bibfnamefont{B.}~\bibnamefont{Eckhardt}},
  \bibinfo{journal}{Physical Review E} \textbf{\bibinfo{volume}{76}},
  \bibinfo{pages}{16301} (\bibinfo{year}{2007}).

\bibitem[{\citenamefont{Malerud et~al.}(1995)\citenamefont{Malerud,
  M{\aa}l{\o}y, and Goldburg}}]{mmg_pof95}
\bibinfo{author}{\bibfnamefont{S.}~\bibnamefont{Malerud}},
  \bibinfo{author}{\bibfnamefont{K.}~\bibnamefont{M{\aa}l{\o}y}},
  \bibnamefont{and} \bibinfo{author}{\bibfnamefont{W.}~\bibnamefont{Goldburg}},
  \bibinfo{journal}{Physics of Fluids} \textbf{\bibinfo{volume}{7}},
  \bibinfo{pages}{1949} (\bibinfo{year}{1995}).

\bibitem[{\citenamefont{Chernykh and Lebedev}(2008)}]{cl_jetp08}
\bibinfo{author}{\bibfnamefont{A.}~\bibnamefont{Chernykh}} \bibnamefont{and}
  \bibinfo{author}{\bibfnamefont{V.}~\bibnamefont{Lebedev}},
  \bibinfo{journal}{JETP Letters} \textbf{\bibinfo{volume}{87}},
  \bibinfo{pages}{682} (\bibinfo{year}{2008}).

\end{thebibliography}
\end{document}